\documentclass[pra,twocolumn]{revtex4-2}
  \usepackage{graphicx}
  \usepackage{amsmath}
  \usepackage{amssymb}
  \usepackage{makeidx}
  \usepackage{amsfonts}
  \usepackage{appendix}
  \usepackage{hyperref}
  \usepackage{mathrsfs}
  \usepackage{appendix}

  \hypersetup
  {
    colorlinks,%
    citecolor=black,%
    linkcolor=black,%
    urlcolor=black,%
  }

  \newcommand{\ket}[1]{\left| #1 \right\rangle}
  \newcommand{\bra}[1]{\left\langle #1 \right|}

\makeindex

\begin{document}

\title {Multimode non-Gaussian secure communication under mode-mismatch}

\author{Soumyakanti Bose}
\email{soumyakanti.bose09@gmail.com}
\affiliation{Department of Physics $\&$ Astronomy, Seoul National University, Gwanak-ro 1, Gwanak-gu, Seoul 08826, Korea}
\author{Hyunseok Jeong}
\email{h.jeong37@gmail.com}
\affiliation{Department of Physics $\&$ Astronomy, Seoul National University, Gwanak-ro 1, Gwanak-gu, Seoul 08826, Korea}

\date{\today}

\begin{abstract}
In this paper, we analyse the role of non-Gaussianity in continuous-variable (CV) quantum key distribution (QKD) with multimode light under mode-mismatch.
We consider entanglement-based protocol with non-Gaussian resources generated by single-photon-subtraction and zero-photon-catalysis on a two-mode squeezed vacuum state (TMSV).
Our results indicate that, compared to the case of TMSV, these non-Gaussian resources reasonably enhances the performance of CV-QKD, even under the effect of noise arising due to mode-mismatch.
To be specific, while in the case of TMSV the maximum transmission distance is limited to $\sim 45$ Km, single-photon subtracted TMSV and zero-photon-catalysed TMSV yield much higher distance of $\sim 70$ Km and $\sim 150$ Km respectively.
However, photon loss as a practical concern in zero-photon-catalysis setup limits the transmission distance for zero-photon-catalysed TMSV to $\sim 35$ Km. 
This makes single-photon-subtraction on TMSV to be the best choice for entanglement-based CV-QKD in obtaining large transmission distance.
Nonetheless, we note that the non-Gaussianity does not improve the robustness of entanglement-based CV-QKD scheme against detection inefficiency.
We believe that our work provides a practical view of implementing CV-QKD with multimode light under realistic conditions.
\end{abstract}
\maketitle

\section{Introduction} 

Encryption and decryption of messages between two distant parties using the rules of quantum mechanics have been a centre of interest over a long period in modern scientific endeavours \cite{qkd_rev_gisin, qkd_rev_pirandola}. 
While classical prescriptions are secured up to the technical limitations in obtaining prime divisors of a large number \cite{rsa}, quantum protocols rely upon the fundamental laws of nature \cite{dvqkd_bennett, dvqkd_ekert, cvqkd_ralph, cvqkd_hillery, cvqkd_cerf, cvqkd_grosshans}. 
Moreover, recent advances indicate the vulnerability of classical cryptography further \cite{prime_factor} and thereby pointing towards the indispensability of quantum cryptography that provides security beyond the scope of classical physics with both asymptotic \cite{dvqkd_sec_shor, dvqkd_sec_acin, dvqkd_sec_pawlowski, cvqkd_asymsec_renner, cvqkd_asymsec_laverrier, cvqkd_asymsec_kogias, cvqkd_asymsec_bradler} and finite resources. \cite{cvqkd_finsec_furrer, cvqkd_finsec_laverrier1, cvqkd_finsec_rupert, cvqkd_finsec_laverrier2, cvqkd_finsec_pap, cvqkd_finsec_lupo}.  

Over last three decades there have been extensive studies on cryptographic aspects of quantum systems, in particular generating/distributing quantum key/password known as quantum key distribution (QKD) \cite{qkd_scarani}.
Communication protocols involving quantum systems could be broadly classified into two groups, discrete variable (DV) QKD \cite{dvqkd_bennett, dvqkd_ekert} and continuous variable (CV) QKD \cite{cvqkd_ralph, cvqkd_hillery, cvqkd_cerf, cvqkd_grosshans}. 
While DV-QKD requires expensive single photon sources, CV-QKD protocols are more readily implementable within the current technology.
Nonetheless, CV protocols are proved to be unconditionally secure against the most general collective attack as well as have been experimentally implemented \cite{cvqkd_exp_grosshans, cvqkd_exp_lodewyck, cvqkd_exp_jouguet, cvqkd_exp_wang, cvqkd_exp_pirandola, cvqkd_exp_jacobsen, cvqkd_exp_huang1, cvqkd_exp_huang2}.

Although quantum optical entangled states with low energy are the ideal choices for performing QKD, it is always challenging to control and manipulate such microscopic systems in practice.
On the other hand, classical light beams which are, in general multimode and bright (intense), easy to operate; however, are devoid of quantum characters.
This often sets a trade-off between quantumness and macroscopicity of the physical systems \cite{macroscopicity_lee}.
In recent times, there have been numerous findings revealing interesting quantum features of such multimode systems \cite{macroquant_hald, macroquant_fernholz, macroquant_iskhakov1, macroquant_toth, macroquant_iskhakov2, macroquant_benbood, macroquant_vasilakis} as well as QKD with them \cite{cvqkdmacro_multient_usenko} and bright light \cite{cvqkdmacro_brightsq_usenko, cvqkdmacro_brightcoh_kovalenko} at the cost of reduced key length.
Although these results promise a practical resolution of controlling sensitive microscopic systems, they are primarily restricted to Gaussian premises only.

On the other hand, over last decade, authors have pointed out the efficiency of various non-gaussian operations in CV-QKD \cite{ngcvqkd_ebsub_huang, ngcvqkd_pmversub_li, ngcvqkd_ebsub_guo, ngcvqkd_mdisub_ma, ngcvqkd_mdiversub_zhao, ngcvqkd_srcat_ye, ngcvqkd_ebcatsub_guo, ngcvqkd_mdisub_kumar, ngcvqkd_ebcatsub_hu, ngcvqkd_mdicat_ye}.
To be particular, non-Gaussianity induced by photon-subtraction \cite{ngcvqkd_ebsub_huang, ngcvqkd_ebsub_guo, ngcvqkd_mdisub_ma, ngcvqkd_mdisub_kumar} or photon-catalysis \cite{ngcvqkd_ebcatsub_guo, ngcvqkd_ebcatsub_hu, ngcvqkd_mdicat_ye} enhances the distance between the parties as well as provides robustness against the detector inefficiency.
However, it remains an open concern whether such non-Gaussian operations have any practical impact on the macroscopic optical systems that play an important role in quantum information processing with optical resources \cite{cvqip_andersen}.
It becomes quite interesting to analyse such non-Gaussian operations in the context of CV-QKD with multimode light.

In the current paper, we analyze QKD with multimode non-Gaussian light under mode-mismatch between the source and the detectors.
We consider entanglement-based  protocol with no-switching assumption \cite{noswitch_weedbrook} as it yields more distance \cite{noswitch_qkd_zhang}. 
In this protocol, two parties generate key by performing heterodyne measurements (measuring both quadrature), instead of homodyne (measuring only one of the quadrature), on the shared entangled state of light.
Although there are many ways of introducing non-Gaussianity on a two-mode squeezed vacuum state (TMSV),  here we consider only single-photon-subtraction and zero-photon-catalysis (i.e., no photon detection at the ancillary mode) as they appear to yield better results \cite{ngcvqkd_mdi_zpcopt_singh}.

We show that both single-photon-subtraction and zero-photon-catalysis enhance the transmission distance considerably, compared to the Gaussian case (TMSV) at all strengths of noise due to mode-mismatch.
In particular, zero-photon-catalysed  TMSV yields the maximum transmission distance of $\sim 150$ Km.
However, considering the effect of photon loss which is a very natural phenomenon occurring in zero-photon-catalysis, we find that single-photon-subtracted TMSV appears to be the better non-Gaussian resource in entanglement-based CV-QKD that yields a maximum distance of $\sim 70$ Km.

Current article is organized as follows.
In Sec. \ref{sec:basic_concept} we introduce the basic notions.
We briefly discuss the multimode homodyne detection with mode-mismatch followed by the channel parameters and derivation of keyrate.
Sec. \ref{sec:qkd_gauss} contains analysis on TMSV.
After briefly describing the constraint on entanglement due to mode-mismatch we present results on keyrate.
In Sec. \ref{sec:mhd_qkd_nongauss} we discuss our simulation results on keyrate for single-photon-subtracted TMSV and zero-photon-catalysed TMSV.
Finally, in Sec. \ref{sec:discussion} we summarize our observations.

\section{Basic Concepts}
\label{sec:basic_concept}

\subsection{Multimode quadrature measurement under mode-mismatch}

\begin{figure}[h]
    \centering
    \includegraphics[scale=0.8]{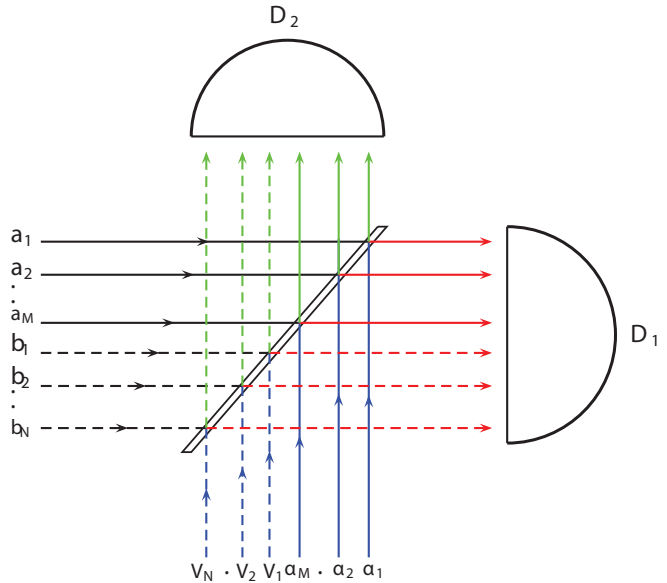}
    \caption{Schematic of homodyne detection of multimode light with mode mismatch.
    Various lines correspond to (a): matched signal mode (green solid line), (b): unmatched signal mode (blue dotted line), (c): local oscillator (pink solid line) and vacuum mode (black dashed line).
    Detectors are designated as $D_1$ and $D_2$; BS being given by red solid line.}
    \label{fig:mhd}
\end{figure}

Let us first consider the basic outline of homodyne detection of a multimode light where the number of modes at the measuring detector differs from the number of modes generated at the source.
We elaborate it in a simple diagram in Fig.~\ref{fig:mhd}.
Suppose the emitter emits total $M+N$ number of signal modes ($a_{i}$) out of which only $M$ of modes match with the local oscillators ($\alpha_i$) ($i=1,2,..,M$) used for quadrature measurement. 
As a consequence, the rest of the $N$ number of emitted signal modes ($b_{j}$) are mixed with the $N$ number of vacuum modes ($V_j$) ($j=1,2,..,N$) in the beam splitter (BS).
For the sake of simplicity we consider balanced homodyne detection, i.e. we use the BS transmission to be $50\%$.

We also consider that the detectors ($D_1$ and $D_2$) can detect the additional modes with efficiency $\sqrt{\epsilon}$. 
For example, say the detector $D_1$ can register the outgoing matched signal modes ($a_{i}^{'}$) completely and the outgoing additional unmatched signal modes ($b_{j}^{'}$) with probability $\sqrt{\epsilon}$.
Consequently, the average photon number, detected at $D_1$ becomes $n_1=\sum_k a_{k}^{'\dagger}a_{k}^{'} + \epsilon \sum_l b_{l}^{'\dagger}b_{l}^{'}$, where primed operators correspond to the respective output modes of the BS. 
Similarly, at $D_2$ the average photon number is given by $n_2=\sum_k \alpha_{k}^{'*}\alpha_{k}^{'} + \epsilon \sum_l V_{l}^{'\dagger}V_{l}^{'}$.

In a simple and straightforward calculation it could be shown that in the presence of this mode-mismatch the measured quadrature for the signal modes changes as \cite{cvqkdmacro_brightsq_usenko} $R_i\rightarrow R_i + \frac{\epsilon}{\alpha_i} \sum_j \left( b_{j}^{\dagger}V_j + b_{j}V_j^{\dagger} \right)$ leading to the variance $\Delta R_i \rightarrow \Delta R_i + \frac{\epsilon^2}{\alpha^2} \sum_j \langle b_{j}^{\dagger}b_{j} \rangle$.
Since there are $M$ number of matched modes, the normalised variance per mode should be obtained by dividing the measured result by the factor $M$.
Furthermore, for the sake of simplicity we consider that the additional (unmatched) modes are all equally strong, i.e., $\langle b_{j}^{\dagger}b_{j} \rangle=\langle b_{k}^{\dagger}b_{k} \rangle = \bar{n}, \forall j,k$.
As a consequence, the normalized variance becomes \cite{cvqkdmacro_brightsq_usenko}
\begin{equation}
    v_i \rightarrow v_i + \frac{N\epsilon^2}{M\alpha^2} \bar{n} = v_i + \delta (\rm{let}).
\end{equation}

Let us denote this $\delta$ as the mode-mismatch-noise.

\subsection{Channel parameters and Keyrate}

In the entanglement-based scheme, one of the parties (say Alice) generates a two-mode entangled resource and sends one of the modes to a distant party (say Bob) through optical cables which are in general lossy.
The loss of the channel due to transmission is quantified as $T=\frac{1}{2} 10^{-l*L}$ where $l=0.02$ (dB/Km) is the loss per Km and $L$ is the distance between Alice and Bob.
This transmittance through lossy channel leads to the {\em line noise} defined as $\chi_{\rm{line}}=\frac{1-T}{T}$.
On the other hand, the homodyne detectors, used by Alice and Bob for measurement on the shared entangled state to generate the key, are not in general prefect. 
This imperfection in the detectors further lead to {\em homodyne noise} as $\chi_{\rm{homo}}=\frac{1-\eta}{\eta}$, where $\eta$ is the detection efficiency.
Under these assumptions, the total additional noise (due to channel transmission and noisy detectors), introduced to the variance matrix could be written as
\begin{equation}
    \chi_{\rm{tot}}=\chi_{\rm{line}}+\frac{2 \chi_{\rm{homo}}}{T}. 
\end{equation}

Let us consider the variance matrix generated by Alice is given as $V=\begin{pmatrix} V_A & V_C\\
V_C^{\mathsf{T}} & V_B
\end{pmatrix}$, where $V_A$ and $V_B$ correspond to the subsystems of Alice and Bob, while $V_C$ is the correlation between them.
Under the effect of lossy channel transmission and imperfect detectors, the final variance matrix becomes $V^{'}=\begin{pmatrix} V_A^{'} & V_C^{'}\\
V_C^{'\mathsf{T}} & V_B^{'}
\end{pmatrix}
=
\begin{pmatrix} V_A & \sqrt{T}V_C\\
\sqrt{T}V_C^{\mathsf{T}} & T(V_B+\chi_{\rm{tot}}\mathbf{I}_2)
\end{pmatrix}$,
where $\mathbf{I}_2$ is the $2\times 2$ identity matrix.
Here, "$\mathsf{T}$" stands for transposition.

It is natural to think of an adversary, say Eve, trying to hack and obtain information about the communication/quantum state between Alice and Bob.
We assume that Eve can perform independent one-mode collective attacks on the communication channel between ALice and Bob.
In this scenario, the secured raw keyrate is given by \cite{keyrate_devetak}
\begin{equation}
    K=\beta I_{\rm{AB}} - \chi_{\rm{hol}},
    \label{eq_keyrate}
\end{equation}
where $I_{\rm{AB}}$ is the mutual information between Alice and Bob and $\chi_{\rm{hol}}$ is the maximum information available to Eve and is given by the Holevo bound \cite{holevo_bound}.
It may be noted that in the case where $I_{\rm{AB}} \leq \chi_{\rm{hol}}$ or simply $I_{\rm{AB}}=0$, $K$ is trivially zero ($K\geq 0$).
This means that if the channel becomes highly noisy/lossy ($I_{\rm{AB}} \leq \chi_{\rm{hol}}$) or we use uncorrelated data set ($I_{\rm{AB}}=0$), there is no secret key.

Furthermore, we consider the reverse reconciliation (Bob communicates his results with Alice) \cite{revrecon_grosshans} as it offers better keyrate as well as is more robust than the direct reconciliation (Alice communicates her results with Bob) \cite{revrecon_chen}.
As a consequence, $\chi_{\rm{hol}}$ is given by the maximum information bound on Eve due to Bob's data and is denoted as $\chi_{\rm{BE}}$.
Let us now consider the transmitted variance matrix between Alice and Bob, $V^{'}$, to evaluate the keyrate.
It is imperative to note that final secured keyrate is obtained after post-processing and privacy amplification carried out on the raw keyrate \cite{qkd_rev_pirandola}.
In the current paper we focus on the raw keyrate \eqref{eq_keyrate} only.

In any CV-QKD protocol, keyrate is obtained by considering the equivalent {\em Prepare-and-Measure} protocol.
Moreover, we consider the no-switching protocol, i.e., instead of homodyne measurements (measuring either of $x$ and $p$), we consider heterodyne measurement (measuring both $x$ and $p$.)
Consequently, the mutual information between Alice and Bob is given by
\begin{equation}
    I_{\rm{AB}} = \frac{1}{2} \log\left( \frac{V_{A_m}^{'x}}{V_{A_m|B_m}^{'x}}\right) +  \frac{1}{2} \log\left( \frac{V_{A_m}^{'p}}{V_{A_m|B_m}^{'p}} \right),
\end{equation}
i.e., contributions coming from the measurements of both $x$ and $p$ quadrature.
Here, $V_{A_m}^{'\zeta}$ and $V_{A_m|B_m}^{'\zeta}$ ($\zeta=x,p$) are the measured quadrature for Alice's subsystem and Alice's conditional subsystem based on Bob's measurement.
These are mathematically described as $V_{A_m}^{'\zeta}=(V_A^{'\zeta} + 1)/2$ and $V_{A_m|B_m}^{'\zeta} = (V_{A|B}^{'\zeta} + 1)/2$ where
\begin{equation}
    V_{A|B}^{'} = V_{A}^{'} - V_C^{'\mathsf{T}}(V_B^{'}+I)^{-1}V_C^{'}.
\end{equation}

On the other hand, the holevo bound between Bob and Eve is defined as \cite{holevo_bound}
\begin{align}
    \chi_{\rm{BE}} &= S(\rho_{\rm{BE}}) - \int dm_B~P(m_B)~S(\rho_{\rm{BE}}^{m_b})
    \nonumber
    \\
    &= S(\rho_{\rm{AB}}) - S(\rho_{\rm{A|B}}),
\end{align}
where $S(\rho)$ denotes the von-Neumann entropy of the state $\rho$, $m_B$ is the Bob's measurement outcome with probability $P(m_B)$ with $\rho_{\rm{BE}}^{m_B}$ is the Eve's state conditioned on the corresponding Bob's measurement. 
In terms of the total variance matrix between Alice and Bob ($V^{'}$) and the Alice's conditional variance matrix based on Bob's measurement ($V_{A|B}^{'}$), one can easily obtain the holevo bound as 
\begin{align}
    S(\rho_{\rm{AB}}) &= G\left( \frac{\lambda_1-1}{2} \right) + G\left( \frac{\lambda_2-1}{2} \right)
    \nonumber
    \\
    S(\rho_{\rm{A|B}}) &= G\left( \frac{\lambda_3-1}{2} \right),
\end{align}
where $G(x)=(x+1)\log_2(x+1) - x\log_2 x$.
$\lbrace \lambda_1, \lambda_2 \rbrace$ and $\lambda_3$ are the symplectic eigenvalues \cite{gaussent_adesso} of $V^{'}$ and $V_{A|B}^{'}$ respectively.

\section{Key Distribution with multimode Gaussian States: Effect of Mode-mismatch}
\label{sec:qkd_gauss}

Here, we consider a two-mode Gaussian resource - a TMSV state described by the variance matrix $V=
\begin{pmatrix} 
\mathbf{A} &\mathbf{C}\\
\mathbf{C} &\mathbf{B}
\end{pmatrix}$, where $\mathbf{A} = \mathbf{B}=\rm{diag}(\eta,\eta)$ correspond to the individual subsystems and $\mathbf{C}=\rm{diag}(c,-c)$ represents the inter-mode correlation with $\eta=\cosh(2r)$ and $c=\sinh(2r)$.
Due to mode-mismatch the measured quantities for the subsystems ($\eta$) would acquire additional contribution while the inter-mode terms ($c$) would be unaffected.
As a consequence, under multimode-homodyne-detection with mode-mismatch, the variance matrix for the TMSV becomes
\begin{equation}
V=\begin{pmatrix}
\eta +\delta &0 &c &0\\
0 &\eta+\delta &0 &-c\\
c &0 &\eta+\delta &0\\
0 &-c &0 &\eta+\delta
\end{pmatrix},
\label{eq_vm_tmsv_mhd}
\end{equation}
where $\delta=\frac{N\epsilon^2}{M\alpha^2} \bar{n}$.

\subsection{Entanglement vs mode-mismatch}
\label{subsec:gauss_ent_mme}

As is evident from the Eq. \eqref{eq_vm_tmsv_mhd}, mode-mismatch introduces additional gaussian noise to the initial variance matrix of the TMSV.
This leads to the mixedness in the variance matrix as $\det V>1$.
As a consequence, we consider the logarithmic negativity to check for the entanglement.
Logarithmic negativity for a bipartite Gaussian state with variance matrix $V=\begin{pmatrix} \mathbf{A} & \mathbf{C}\\
\mathbf{C} & \mathbf{B}
\end{pmatrix}$ is given in terms of its minimum symplectic eigenvalue (under partial transposition) $l_{\min}$ as \cite{gaussent_adesso} $\mathbf{E_N}=\max \lbrace 0,-\log l_{\min} \rbrace$, where
\begin{equation}
    l_{\min}=\sqrt{\frac{\Delta - \sqrt{\Delta^2-4\det V}}{2}}
\end{equation}
and
 $\Delta=\det A + \det B - 2\det C$. 

Consequently, in the present case of multimode homodyne detection with mode-mismatch, logarithmic negativity for the variance matrix $V$ \eqref{eq_vm_tmsv_mhd} is given by
\begin{equation}
    \mathbf{E_N}=-\frac{1}{2} \log \left(-1+\delta^2+2(\delta+\mu)(\mu-\nu)\right)
    \label{eq_logneg_mhd}
\end{equation}
where $\mu=\cosh r$ and $\nu=\sinh r$. 

It is straightforward to note that the entanglement ($\mathbf{E_N}$) is a strictly decreasing function of the mode-mismatch-noise ($\delta$). 
One can further show that the condition for the inseparability ($\mathbf{E_N}\geq 0$) \cite{gaussent_adesso} for the variance matrix $V$ $\left( \Delta>\det V + 1 \right)$, is given by
\begin{equation}
    \delta<1-\cosh 2r + \sinh 2r,
\end{equation}
For $\delta\geq 1-\cosh 2r + \sinh 2r$, $V$ represents a separable state.
In other words, there is no secret key as $I_{\rm{AB}}$ becomes zero.
Moreover, for $\delta=1$, the state is always separable, i.e., for $\delta=1$ there is no entanglement. 

\subsection{Keyrate analysis}

\begin{figure}
    \centering
    \includegraphics[angle=-90,scale=0.35]{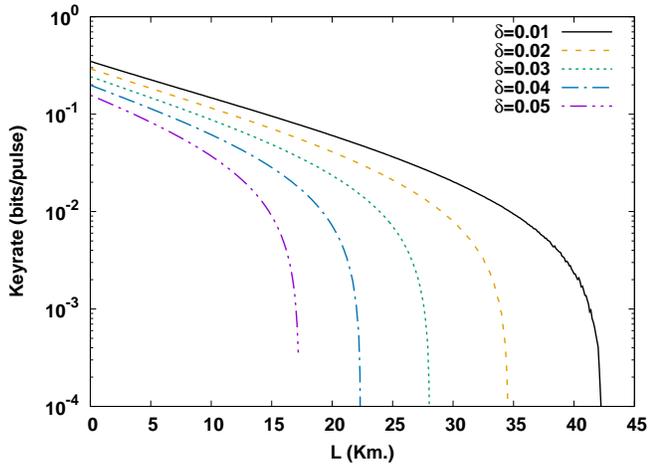}
    \caption{Keyrate vs length for TMSV for different noise parameter $\delta$.
    Other parameters are $T_{\rm{bs}}=0.9$ and $\eta=1.0$.
    Different curves correspond to $\delta=0.01$ (black solid line), $\delta=0.02$ (yellow dashed line), $\delta=0.03$ (green dotted line), $\delta=0.04$ (blue dashed-dotted line) and $\delta=0.05$ (purple dashed-double-dotted line).}
    \label{fig:kvsl_tmsv}
\end{figure}

In figs. \ref{fig:kvsl_tmsv} we plot the dependence of keyrate upon the distance $L$ for TMSV.
The maximum transmission distance for TMSV, i.e., $\sim 42$ km, is obtained with the mode-mismatch-noise $\delta=0.01$.
As the error increases, the transmission distance decreases.

\begin{figure}
    \centering
    \includegraphics[angle=-90,scale=0.35]{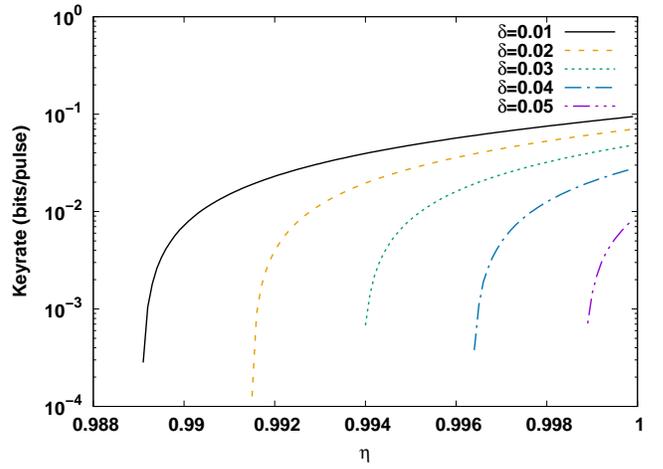}
    \caption{Keyrate vs $\eta$ for TMSV for different $\delta$.
    Other parameters are $T_{\rm{bs}}=0.9$ and $L_{\rm{ab}}=15$ km.
    Different curves correspond to $\delta=0.01$ (black solid line), $\delta=0.02$ (yellow dashed line), $\delta=0.03$ (green dotted line), $\delta=0.04$ (blue dashed-dotted line) and $\delta=0.05$ (purple dashed-double-dotted line).}
    \label{fig:kvseta_tmsv}
\end{figure}

Similarly, in Fig. \ref{fig:kvseta_tmsv} we plot the dependence for keyrate upon detector efficiency for TMSV.
As the mode-mismatch-noise increases, system becomes more sensitive to the detector efficiency. 
It may further be noted that for moderate mode-mismatch ($\delta\geq 0.08$) it is almost impossible to obtain any key irrespective of the detector efficiency.

\section{Key Distribution with non-Gaussian sampling of macroscopic Gaussian States: Mitigating the Effect of Mode-mismatch}
\label{sec:mhd_qkd_nongauss}

In this section, we consider the role of non-Gaussianity to mitigate the effect of the mode-mismatch.
Non-gaussianity generated in various ways such as photon subtraction, addition as well as catalysis plays an important role in enhancing the performance of QKD.
Here, we analyze two of such de-gaussification processes on TMSV, such as single-photon-subtraction and zero-photon-catalysis.
Corresponding states are denoted as single-photon-subtracted TMSV and zero-photon-catalysed TMSV.

\subsection{Linear optical scheme for photon-Subtraction and zero-photon-catalysis}
\label{subsec:scheme_pszpc}

\begin{figure}[h]
    \centering
    \includegraphics[scale=0.8]{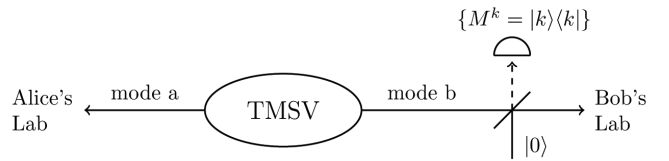}
    \caption{Schematic of photon subtraction/catalysis.
    Detection of $k$-photon in the ancilla leads to $k$-photon subtraction in mode b.
    $k=0$ stands for zero-photon-catalysis.}
    \label{fig:ps_diag}
\end{figure}

Let us first consider the schematic as shown in fig. \ref{fig:ps_diag} for a unified view of photon-subtraction and zero-photon-catalysis.
The process of photon-subtraction/catalysis on TMSV could be explained as follows,

{\em Step 1}: First we pass one of the modes of TMSV through a BS with transmittance $T_{\rm{bs}}$ while the other input of the BS is left at vacuum.

{\em Step 2}: On the output ancila mode we detect $k$ number of photons through a photon number resolving detector for which the measurement operators are given by $\lbrace M^k; M^k=\ket k\bra k \rbrace$.
This process is described in the Hilbert space description as considering the overlap between the output state and the number state $k$ in the ancila mode.

The process of photon subtraction is probabilistic where the probability of $k$ photon subtraction is given by \cite{ngcvqkd_mdisub_ma} $P(k)=\frac{1}{\mu^2} \frac{\tau^{2k}(1-T_{\rm{bs}})^k}{(1-\tau^2 T_{\rm{bs}})^{k+1}}$, where $\mu=\cosh r$ and $\tau=\tanh r$.
It could be noted that $k=0$ corresponds to the case of zero-photon catalysis on TMSV.
The variance matrix for $k$-photon subtracted TMSV is given by \cite{ngcvqkd_mdisub_ma} $V^{(k)}=\begin{pmatrix} x^{(k)}\mathbf{I} & z^{(k)} \sigma_3 \\
z^{(k)} \sigma_3 & y^{(k)}\mathbf{I}
\end{pmatrix}$,
where $\mathbf{I}$ is the $2\times 2$ identity matrix, $\sigma_3=\rm{diag}(1,-1)$ is the pauli matrix, $x^{(k)}=\frac{2(1+k)}{1-\tau^2 T_{\rm{bs}}}-1$, $y^{(k)}=\frac{2(1+k \tau^2 T_{\rm{bs}})}{1-\tau^2 T_{\rm{bs}}}-1$ and $z^{(k)}=\frac{2\sqrt{T_{\rm{bs}}}\tau(1+k)}{1-\tau^2 T_{\rm{bs}}}$.
It may be noted that due mode-mismatch, both $x^{(k)}$ and $y^{(k)}$ attains additional contribution of $\delta$ similar to the case of TMSV \eqref{eq_vm_tmsv_mhd}.

It is worth mentioning here that in the previous studies \cite{ngcvqkd_ebsub_huang, ngcvqkd_pmversub_li, ngcvqkd_ebsub_guo, ngcvqkd_mdisub_ma, ngcvqkd_mdiversub_zhao, ngcvqkd_srcat_ye, ngcvqkd_ebcatsub_guo, ngcvqkd_mdisub_kumar, ngcvqkd_ebcatsub_hu, ngcvqkd_mdicat_ye}, authors have considered the keyrate expression $K=P_{\rm{res}}\left( \beta I_{\rm{AB}} - \chi_{\rm{hol}} \right)$, where $P_{\rm{res}}$ is the probability of generating the specific non-gaussian resource.
However, the processes for generating the resources are {\em offline processes}, i.e., the rest of the key distribution protocol are executed only after we can successfully generate the resources.
From this perspective, every successful communication between Alice and Bob must take into account the specific non-gaussian resource not the original gaussian TMSV. 
As a consequence, in the current work we have ignored the probability factor and used the keyrate expression $K=\beta I_{\rm{AB}} - \chi_{\rm{hol}}$ \eqref{eq_keyrate}.
Next, we analyze keyrate with the distance between Alice and Bob ($L$) as well as the detector efficiency ($\eta$) for single-photon-subtracted TMSV and zero-photon-catalysed TMSV.

\subsection{Keyrate analysis for single-photon-subtracted TMSV:}
\label{subsec:keyrate_1pstmsv}

\begin{figure}
    \centering
    \includegraphics[angle=-90,scale=0.35]{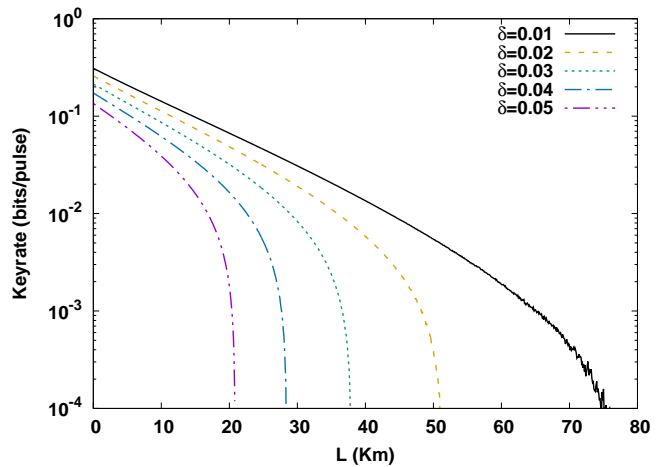}
    \caption{Keyrate vs length for 1PSTMSV for different noise parameter $\delta$.
    Other parameters are $T_{\rm{bs}}=0.9$ and $\eta=1.0$.
    Different curves correspond to $\delta=0.01$ (black solid line), $\delta=0.02$ (yellow dashed line), $\delta=0.03$ (green dotted line), $\delta=0.04$ (blue dashed-dotted line) and $\delta=0.05$ (purple dashed-double-dotted line).}
    \label{fig:kvsl_1pstmsv}
\end{figure}

\begin{figure}
    \centering
    \includegraphics[angle=-90,scale=0.35]{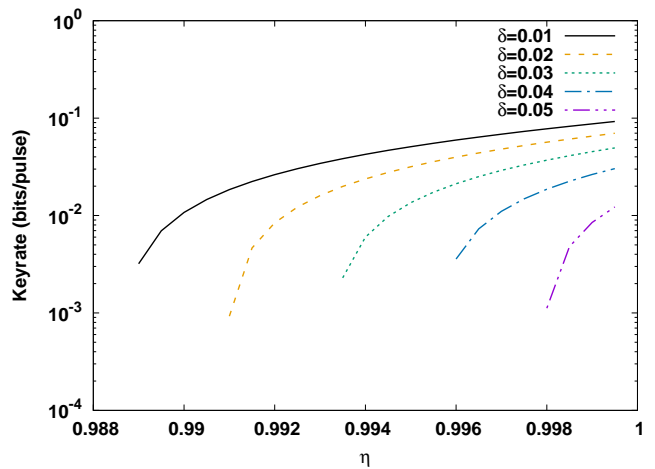}
    \caption{Keyrate vs $\eta$ for 1PSTMSV for different $\delta$.
    Other parameters are $T_{\rm{bs}}=0.9$ and $L_{\rm{ab}}=15$ km.
    Different curves correspond to $\delta=0.01$ (black solid line), $\delta=0.02$ (yellow dashed line), $\delta=0.03$ (green dotted line), $\delta=0.04$ (blue dashed-dotted line) and $\delta=0.05$ (purple dashed-double-dotted line).}
    \label{fig:kvseta_1pstmsv}
\end{figure}

In Figs. \ref{fig:kvsl_1pstmsv} and \ref{fig:kvseta_1pstmsv} we plot the dependence of keyrate ($K$) on the transmission distance ($L$) and the detector efficiency ($\eta$) for single-photon-subtracted TMSV, respectively.
Evidently, with the minimum additional noise ($\delta=0.01$) the maximum distance is $\sim 75$ Km, almost $30$ km more than that for TMSV.
However, photon subtraction doesn't improve the robustness against detector inefficiency significantly.
As it is evident from Fig. \ref{fig:kvseta_1pstmsv}, the lowest possible detection efficiency to obtain keyrate is $\sim 0.988$.
Nonetheless, the overall of pattern of keyrate w.r.t. the mme remains same.
As the noise ($\delta$) increases performance drops.

\subsection{Keyrate analysis for zero-photon-catalysed TMSV}
\label{subsec:keyrate_zpctmsv}

\begin{figure}
    \centering
    \includegraphics[angle=-90,scale=0.35]{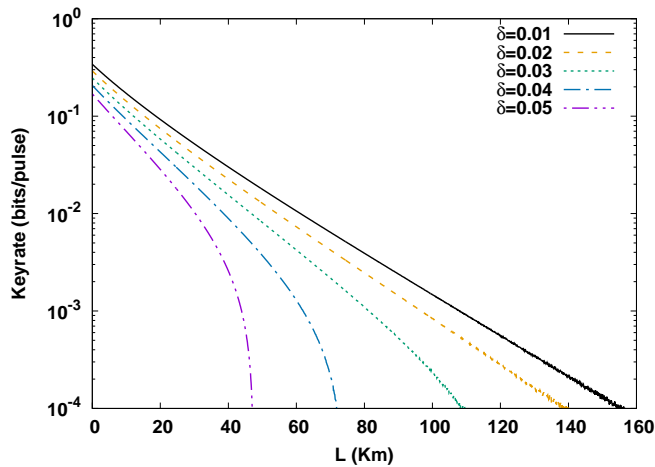}
    \caption{Keyrate vs length for ZPCTMSV for different noise parameter $\delta$.
    Other parameters are $T_{\rm{bs}}=0.9$ and $\eta=1.0$.
    Different curves correspond to $\delta=0.01$ (black solid line), $\delta=0.02$ (yellow dashed line), $\delta=0.03$ (green dotted line), $\delta=0.04$ (blue dashed-dotted line) and $\delta=0.05$ (purple dashed-double-dotted line).}
    \label{fig:kvsl_zpctmsv}
\end{figure}

\begin{figure}
    \centering
    \includegraphics[angle=-90,scale=0.35]{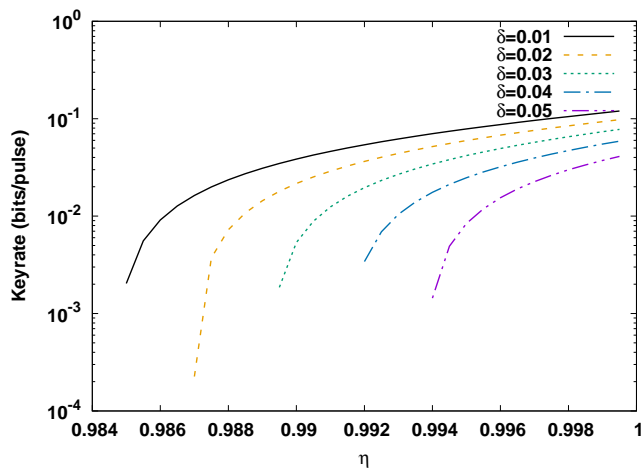}
    \caption{Keyrate vs $\eta$ for ZPCTMSV for different $\delta$.
    Other parameters are $T_{\rm{bs}}=0.9$ and $L_{\rm{ab}}=15$ km.
    Different curves correspond to $\delta=0.01$ (black solid line), $\delta=0.02$ (yellow dashed line), $\delta=0.03$ (green dotted line), $\delta=0.04$ (blue dashed-dotted line) and $\delta=0.05$ (purple dashed-double-dotted line).}
    \label{fig:kvseta_zpctmsv}
\end{figure}

In Figs. \ref{fig:kvsl_zpctmsv} and \ref{fig:kvseta_zpctmsv} we plot the keyrate against transmission distance ($L$) and detector efficiency ($\eta$) respectively for zero-photon-catalysed TMSV.
As it is evident, photon-catalysis yields the maximum transmission distance. 
To be specific, with mode-mismatch-noise $\delta=0.01$, the maximum transmission distance is $\sim 150$ km - almost $100$ Km more than TMSV and $80$ Km more than single-photon-subtracted TMSV.
However, it fails to to improve robustness against the detector inefficiency, as compared to both TMSV and single-photon-subtracted TMSV.
The lowest possible detection efficiency at which QKD could be operated is $\sim 0.984$ (Fig. \ref{fig:kvseta_zpctmsv}).

\subsubsection{Zero-photon-catalysis and photon loss}
\label{subsubsec:zpc_photonloss}

Before we conclude, it is important to look at the practical concern of photon loss in generating zero-photon-catalysed TMSV.
As is described in the Sec. \ref{subsec:scheme_pszpc}, a zero-photon-catalysed TMSV is generated when the detector at the outgoing ancilla mode registers no photon or, in other words, the detector does not click.
Under ideal condition or say perfect experimental setup, no click in the detector  means collapse of the state in the $\ket 0\bra 0$ state yielding zero-photon-catalysis.
However, in reality, there is an intrinsic technical issue in generating zero-photon-catalysed TMSV using this method, as discussed below.

The {\em no click} condition can appear in characteristically two different situations - when the photon in the outgoing ancilla mode {\bf (a)} is collapsed in the $\ket 0\bra 0$ state or {\bf (b)} is lost. 
As a consequence, the effective variance matrix between Alice and Bob becomes,
\begin{equation}
    V_{\rm{AB}}=p V_{\rm{lost}} + (1-p) V_{\rm{zpc}},
    \label{eq_vm_zpcloss}
\end{equation}
where $p$ is the probability of losing the outgoing ancilla photon and $V_{\rm{zpc}}$ is the variance matrix for zero-photon-catalysed TMSV. 
The variance matrix for the photon loss case ($V_{\rm{lost}}$) is given by $V_{\rm{lost}} = 
\begin{pmatrix}
\cosh 2r \mathbf{I} &\mathbf{0}\\
\mathbf{0} &\mathbf{I}
\end{pmatrix}$,
where $\mathbf{I}$ and $\mathbf{0}$ are the $2\times 2$ identity and null matrices.

\begin{figure}
    \centering
    \includegraphics[angle=-90,scale=0.35]{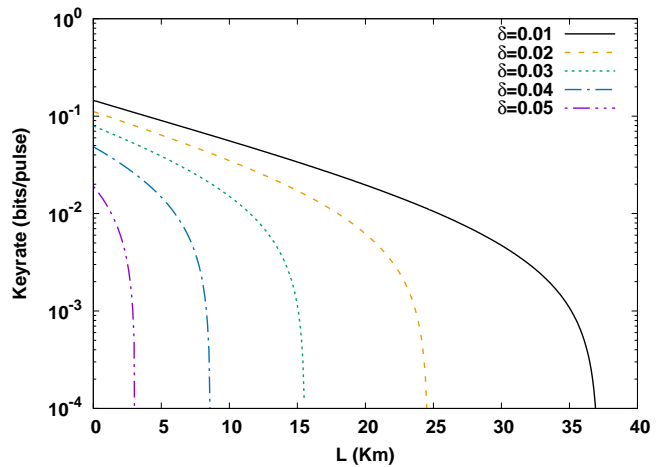}
    \caption{Keyrate vs transmission distance for ZPCTMSV under the effect of photon loss for different $\delta$.
    The photon loss probability is kept very low as $p=0.002$.
    Different curves correspond to $\delta=0.01$ (black solid line), $\delta=0.02$ (yellow dashed line), $\delta=0.03$ (green dotted line), $\delta=0.04$ (blue dashed-dotted line) and $\delta=0.05$ (purple dashed-double-dotted line).}
    \label{fig:kvsl_zpctmsv_loss}
\end{figure}

To illustrate the effect of photon loss in the generation of zero-photon-catalysed TMSV, in Fig. \ref{fig:kvsl_zpctmsv_loss} we plot the keyrate vs length for a very low probability of losing photon $p=0.002$.
As compared to the no photon loss case (Fig. \ref{fig:kvsl_zpctmsv}), even for such a small probability, the transmission distance reduces significantly - from $\sim 150$ Km to $\sim 35$ Km.
This may be interpreted as follows. 
The variance matrix for the photon loss case \eqref{eq_vm_zpcloss} essentially represents and gaussian lossy channel.
Now, at the operating parameter region $\cosh 2r=50$, the additional noise $p V_{\rm{lost}}$ in the variance matrix ($V_{\rm{AB}}$) is significantly high to reduce the effective correlation and thus the keyrate between ALice and Bob.

\section{Discussion}
\label{sec:discussion}

In this paper, we have analysed the role of non-Gaussianity in CV-QKD with multimode light under mode-mismatch.
We have shown that both the non-Gaussian operations, single-photon subtraction as well as zero-photon-catalysis, are helpful in mitigating the noise due to the presence of unmatched modes and improve the overall performance reasonably, as compared to the Gaussian case.
Under ideal state generation setup, zero-photon-catalysis, in comparison to single-photon-subtraction, offers the optimal results. 
It offers the maximum distance between the parties around $150$ km while maintaining the keyrate above $10^{-4}$ bits/pulse in the realistic situation.

However, in consideration of the photon loss, a major concern of zero-photon-catalysis setup, the best performance of zero-photon-catalysed TMSV is no longer true. 
It appears that under the effect of photon loss, zero-photon-catalysed TMSV offers much less transmission distance $\sim 35$ Km - that is less than the case with Gaussian TMSV.
On the other hand, such a situation does not arise in the case of single-photon-catalysed TMSV as the QKD protocol proceeds only after the photon is detected in the detector (which may be made more precise by tuning $T_{\rm{bs}}$).
Thus, in comparison to zero-photon catalysis, single-photon subtraction offers the optimal practical solution for obtaining larger transmission distance in non-Gaussian QKD in contrast to the earlier results \cite{ngcvqkd_ebcatsub_guo, ngcvqkd_ebcatsub_hu, ngcvqkd_mdicat_ye, ngcvqkd_mdi_zpcopt_singh}.
Nonetheless, it must be noted that both zero-photon-catalysed TMSV and single-photon-catalysed TMSV fail to improve the robustness of the entanglement-based CV-QKD against the detector inefficiency.

For the keyrate analysis, we have considered a simple model for the noise factors that are present in the communication channel.
However, in reality, there may be additional factor such as gain of the homodyne measurement, electronic noise of the detector {\em etc.} that further reduces the keyrate as well as the maximum transmission distance \cite{ngcvqkd_mdisub_ma}.
It may be noted that earlier, analysis of CV-QKD with multimode Gaussian light was centred around mitigating the effect of mode-mismatch by considering bright light \cite{cvqkdmacro_brightsq_usenko} as well as rearranging the multiple modes \cite{cvqkdmacro_multient_usenko}.
Moreover, here we consider key distribution protocol with both the quadrature unlike the earlier works \cite{cvqkdmacro_multient_usenko, cvqkdmacro_brightsq_usenko, cvqkdmacro_brightcoh_kovalenko} where only one of the quadrature was measured.
Present works offers a different perspective in terms of enhancement in the performance with the use of non-Gaussianity as resource.
We have considered single-photon-subtracted TMSV and zero-photon-catalysed TMSV as the two non-gaussian resources, in comparison to gaussian TMSV.

One can easily extend the our analysis on the role of non-Gaussianity in CV-QKD with different kind of noise such as state-preparation-error \cite{ngcvqkd_preperr_derkach, ngcvqkd_preperr_zhang, ngcvqkd_preperr_jain} where the initial state suffers from side-channel loss prior to modulation.
Nonetheless, compared to the entanglement-based protocol, one may further go for more theoretically motivated model such as measurement-device-independent protocol which are more promising from the point of view of ensuring security \cite{ngcvqkd_mdisub_ma, ngcvqkd_mdiversub_zhao, ngcvqkd_mdisub_kumar, ngcvqkd_mdicat_ye, ngcvqkd_mdizpc_water_wang, ngcvqkd_mdi_zpcopt_singh}.
Nonetheless, it will be interesting to analyse the effect of post-selection rather than actual state generation in non-Gaussian CV-QKD \cite{ngcvqkd_pmzpc_zhong} where one can avoid the issues related to the generation of non-Gaussian resources.
In view of the recent advances on the non-Gaussian operations, we believe our work provides a realistic scheme to implement non-Gaussian CV-QKD in a metropolitan area within the current boundary of technology.

\acknowledgments

This work was supported by the National Research Foundation of Korea (NRF) grants funded by the Korea
government (Grant Nos. NRF-2020R1A2C1008609, NRF-2020K2A9A1A06102946, NRF-2019R1A6A1A10073437 and NRF-2022M3E4A1076099) via the Institute of Applied Physics at Seoul National University, and by the Institute of Information $\&$ Communications Technology Planning $\&$ Evaluation (IITP) grant funded by the Korea government (MSIT) (IITP-2021-0-01059 and IITP-2022-2020-0-01606).

\end{document}